# The Dynamics of Functional Brain Networks: Integrated Network States during Cognitive Function

J. M. Shine[1,2*], P. G. Bissett[1], P. T. Bell[3], O. Koyejo[1], J. H. Balsters[4], K. J. Gorgolewski[1], C. A. Moodie[1], and R. A. Poldrack[1]

1 – Department of Psychology, Stanford University, Stanford, CA, USA
2 – Neuroscience Research Australia, The University of New South Wales, Sydney, NSW, Australia
3 – Queensland Brain Institute, The University of Queensland, Brisbane, QLD, Australia
4 – Department of Health Sciences and Technology, Neural Control of Movement Laboratory, ETH Zurich, Switzerland

* – Corresponding author. Email: macshine@stanford.edu

**Abstract**

Higher brain function relies upon the ability to flexibly integrate information across specialized communities of brain regions, however it is unclear how this mechanism manifests over time. In this study, we use time-resolved network analysis of functional magnetic resonance imaging data to demonstrate that the human brain traverses between two functional states that maximize either segregation into tight-knit communities or integration across otherwise disparate neural regions. The integrated state enables faster and more accurate performance on a cognitive task, and is associated with dilations in pupil diameter, suggesting that ascending neuromodulatory systems may govern the transition between these alternative modes of brain function. Our data confirm a direct link between cognitive performance and the dynamic reorganization of the network structure of the brain.

Within the brain, a highly dynamic functional landscape unfolds on a relatively fixed structural scaffold (Deco et al., 2015; Shen et al., 2015) in which the emergence of momentary neural coalitions forms the basis for complex cognitive functions (Bassett et al., 2015; Cole et al., 2014), learning (Bassett et al., 2011) and consciousness (Barttfeld et al., 2015). This view of brain function highlights the role of individual brain regions within the context of a broader neural network (Bullmore and Sporns, 2012), and recent work has noted the importance of time in understanding the functional relevance of alterations in this network structure under different behavioral conditions (Varela et al., 2001).

Time-resolved analyses of functional neuroimaging data provide a unique opportunity to examine these time-varying reconfigurations in global network structure. These experiments provide a sensitive method for identifying time-sensitive shifts in inter-areal synchrony, which has been proposed as a key mechanism for effective communication between distant neural regions (Fries, 2015). To this end, recent experiments using functional MRI data have demonstrated that global brain signals transition between states of high and low connectivity strength over time (Zalesky et al., 2014) and that these fluctuations are related to coordinated patterns of network topology (Betzel et al., 2015), however the direct psychological relevance of these fluctuations in network topology remain poorly understood.

In the present work, we show that dynamic fluctuations in network structure relate to ongoing cognitive function, and further demonstrate a relation between these fluctuations and integration within a network of frontoparietal and subcortical regions that track with the ascending neuromodulatory system of the brain, as characterized using pupillometry (Joshi et al., 2016). Together, the results of our experiments provide mechanistic evidence to support the role of global network integration in effective cognitive performance.

## Results

### Fluctuations in Network Cartography

To elucidate fluctuations in the network structure of the brain over time, we computed a windowed estimate of functional connectivity (Shine et al., 2015) from a cohort of 92 unrelated subjects obtained from the Human Connectome Project (HCP; see *Materials and Methods*; Smith et al., 2013). After identifying the community structure of the brain's functional connectivity network (Rubinov and Sporns, 2010), we estimated the importance of each region for maintaining this network structure (which changed within each temporal window) by calculating its connectivity both within ($W_T$) and between ($B_T$) each community (see *Materials and Methods*; Guimerà and Nunes Amaral, 2005; Sporns and Betzel, 2015). While previous studies have clustered these metrics at the regional level using pre-defined cartographic boundaries (Guimerà and Nunes Amaral, 2005; Mattar et al., 2015), we hypothesized that the brain should fluctuate as a whole between cartographic extremes that were characterized by either segregation (i.e. the extent to which communication occurs primarily within tight-knit communities of regions) or integration (i.e. the degree of communication between distinct regions; Deco et al., 2015), which might otherwise be obscured by reduction into classes defined by these arbitrary cartographic boundaries.

To test this hypothesis in the resting state, we created a novel analysis technique to assess the temporal classification into two states without requiring the grouping of each region into a pre-defined cartographic class (Guimerà and Nunes Amaral, 2005) which we refer to here as the "cartographic profile".

Subject-level *k*-means clustering of these full profiles across time (*k* = 2, with stable clustering at higher values of *k*; see *Materials and Methods* and Figure S1) identified two states that were characterized by either Integration or Segregation (Figure 1a). The resting brain explored a dynamical repertoire within this topological space, fluctuating aperiodically between the integrated and segregated temporal states, with the majority of time spent in the integrated state (70.32 ± 1.4% of rest session). Although the majority of the group-level fluctuations occurred in inter-modular connectivity (i.e. $B_T$ values transitioned between high and low states *en masse*), we also observed window-to-window fluctuations in intra-modular connectivity ($W_T$) within individual parcels (see Videos 1 and 3 at http://github.com/macshine/coupling demonstration of the fluctuations of the cartographic profile over time).

The two states also showed differential patterns of regional inter-modular connectivity (Figures 1c and 2d), with the integrated state characterized by a global increase in inter-modular communication across the brain (FDR $p < 0.05$ for all 375 individual parcels). This was also reflected in graph-theoretic measures of network-wide integration (global efficiency; Bullmore and Sporns, 2012) and segregation (modularity; Sporns and Betzel, 2015): temporal windows associated with the segregated state had significantly elevated modularity ($Q_S$ = 0.55 ± 0.1 *vs.* $Q_I$ = 0.42 ± 0.2; Cohen's d = 0.9; $p = 10^{-11}$; Figure S2) whereas those associated with the integrated state had greater global efficiency ($E_S$ = 0.18 ± 0.03 *vs.* $E_I$ = 0.24 ± 0.05; Cohen's d = 1.5; $p = 10^{-8}$; Figure S1). The shift towards integration was most prominent in sensory and attentional networks (Figure 1d; FDR $p < 0.05$), whereas the segregated state was associated with relatively higher participation within regions in the default mode network, suggesting that the cartographic profile may reflect changes in the engagement of attention and cognition over time (Corbetta and Shulman, 2002). Importantly, the fluctuations in global network topology occurred independently of the mean framewise displacement in each TR (mean r = 0.01 ± 0.01), nuisance signals from cerebrospinal fluid and deep cerebral white matter (mean r = -0.02 ± 0.01) and of the number of modules estimated within each temporal window (mean r = 0.03 ± 0.10).

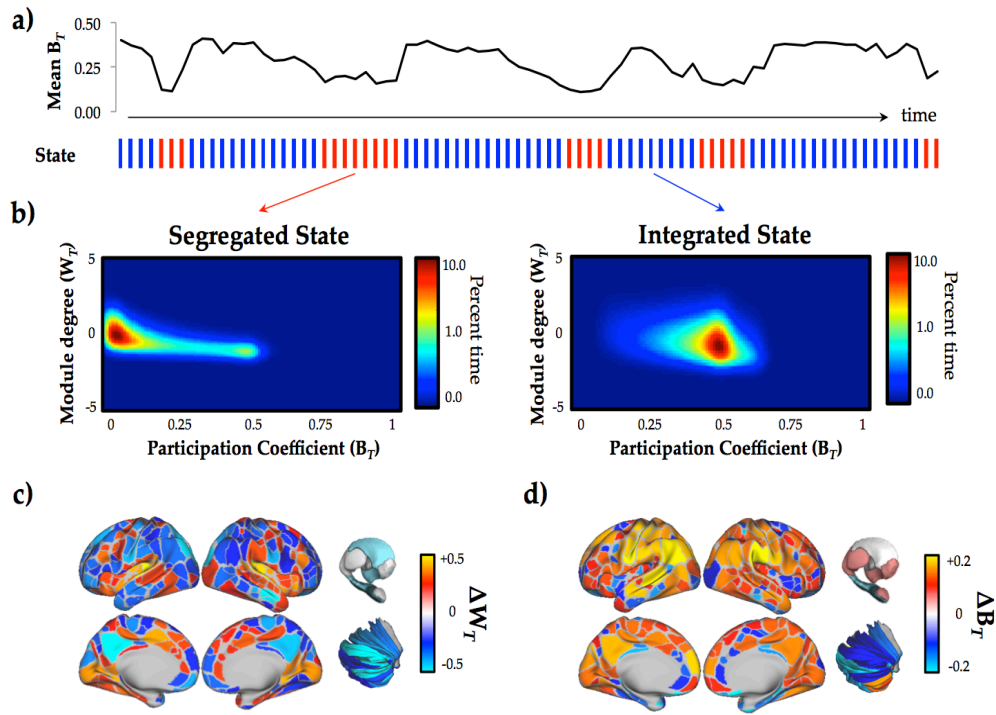

**Figure 1: Dynamic fluctuations in cartography:** a) upper: a representative time series of the mean $B_T$ for a single individual from the Discovery cohort (HCP #100307); lower: each temporal window was partitioned into one of two topological 'states' using *k*-means clustering (red: 'Segregated' and blue: 'Integrated'); b) the mean cartographic profile of both the Segregated and Integrated states (HCP Discovery cohort; n = 92); c) regions with greater $W_T$ in the Integrated > Segregated state; and d) regions with greater $B_T$ in the Integrated > Segregated state.

**Task-based Alterations in the Cartographic Profile**

We next examined whether the balance between network integration and segregation tracked with ongoing cognitive function using data from a cognitively-challenging "N-back" task (Barch et al., 2013). We observed a strong correlation between fluctuations in cartography across all parcels and the blocks of the experimental task (group mean Pearson's r = 0.521; $R^2$ = 0.27; p = $10^{-10}$; Figure 2a & Video 2 [http://github.com/macshine/coupling]), as well as a distinct alteration in the cartographic profile when compared to the resting state (Figure 2b). In addition, the extent of integration remained correlated with the task regressor even after controlling for the global signal (mean r = 0.452 ± 0.21; p = $10^{-10}$) and the mean time-resolved connectivity across all parcels (mean r = 0.393 ± 0.14; p = $10^{-9}$), suggesting that the fluctuations in topology were not driven purely by a constraint in signal processing imposed by the task structure.

Together, these results suggest that the brain transitions into a state of higher global integration in order to meet extrinsic task demands. Indeed, all of the 375 regions showed a significant shift towards greater inter-modular connectivity ($B_T$) during the N-back task when compared to the resting state (FDR p < 0.05 for all 375 regions). Despite this global shift towards integration, the effect was most pronounced within frontoparietal, default mode and subcortical regions (Figure 2c), many of which have been previously identified as belonging to a 'rich club' of densely-interconnected, high degree 'hub' nodes that are critical for the resilience and stability of the global brain network (van den Heuvel and Sporns, 2013). Importantly, the involvement of these highly interconnected regions during the task would likely facilitate effective communication between specialist regions that would otherwise remain isolated, thus affording a larger repertoire of potential responses to deal with the cognitive challenges of the task.

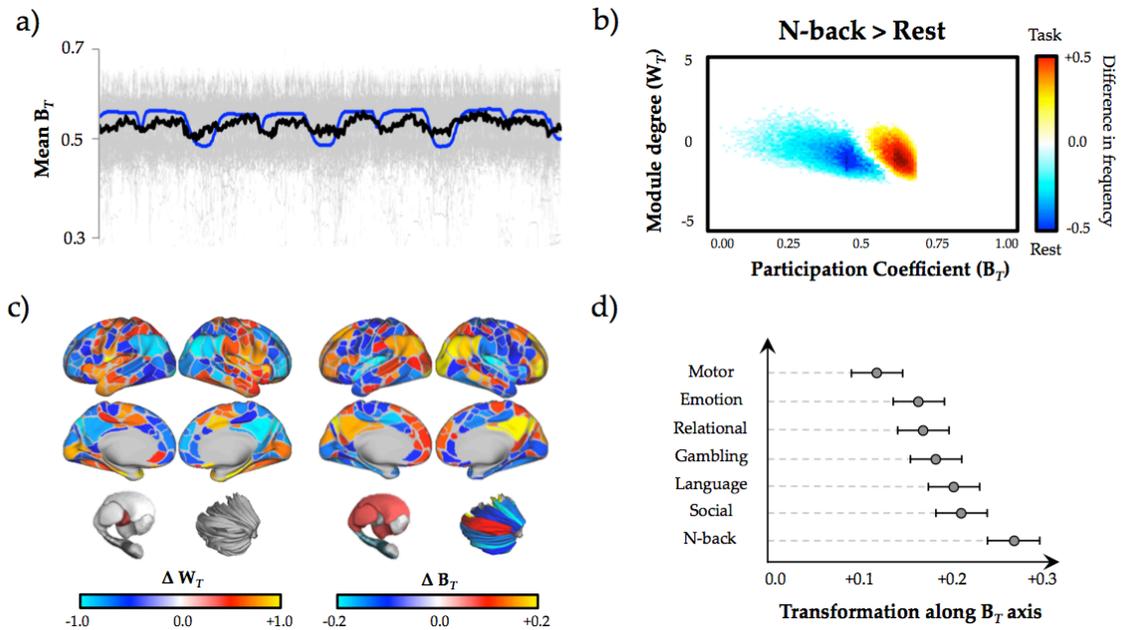

**Figure 2: Alteration of cartographic profile during task performance:** a) time series plot demonstrating the close temporal relationship between mean $B_T$ across 100 subjects (thick black line; individual subject data plotted in grey) and task-block regressors (blue line) – Pearson's correlation between regressor and group mean $B_T$: r = 0.521; b) regions of the 2-dimensional joint histogram that were significantly different between N-back task blocks and the resting state (paired-samples t-test) – colored points indicate regions that survived false discovery correction (FDR p < 0.05): red/yellow – increased frequency during N-back task blocks; blue/light blue – increased frequency during resting state (FDR p < 0.05); c) surface projections of parcels associated with higher $W_T$ (left) or $B_T$ (right) during the N-back task, when compared the resting state – frontoparietal and subcortical 'hub' regions showed elevated $B_T$ during task, whereas $W_T$ was elevated in primary systems and decreased in default mode regions; d) a plot quantifying the shift away from the cartographic profile in the resting state (along the between-module ($B_T$) connectivity axis) across the six tasks in the HCP dataset.

To determine whether network topology was sensitive to task complexity, we calculated the cartographic profile in the remaining six tasks from the HCP in the same cohort of 92 subjects (Barch et al., 2013). While the performance of each task also led to an increase in global integration relative to rest, the effect was less pronounced than the movement observed in the N-back task, particularly when compared to the relatively simple Motor task (88.8% of parcels showed higher $B_T$ in the N-back task; FDR p < 0.05). This effect was quantified by estimating the affine transformation required to align each subjects resting cartographic profile with their profile during each task (transformation along the $B_T$ axis relative to rest; Figure 2d). Together, these results suggest that the extent of reconfiguration during task performance varied as a function of cognitive demands.

**Investigating the Relationship Between Cartography and Behavior**

Based on these findings, we predicted that a more globally integrated network architecture would give rise to faster, more effective information processing during task performance. To test this hypothesis, we fit a drift diffusion model to each subject's behavior (response time distributions and accuracy) on the more cognitively challenging 2-back trials within the N-back task using the EZ-diffusion model (22) (Figure 3a). The diffusion model provides a decomposition of behavioral performance into cognitively-relevant latent variables representing the speed and accuracy of information processing (drift rate; '$v$'), the speed of perceptual and motor processes not directly related to the decision process (non-decision time; '$t$') and a flexible measure of response caution (boundary separation; '$a$') (Ratcliff, 1978). Theoretically, faster progression throughout all stages of information processing from perception through action should be reflected in a positive relationship between global integration and both faster drift rate and shorter non-decision time, whereas integration should be independent of the boundary parameter.

We compared these model parameters to the mean N-back cartographic profile across the Discovery cohort (Figure 3a). The extent of global network integration in the cartographic profile was positively correlated with drift rate (Figure 3b), inversely correlated with non-decision time (Figure 3c), and had no relationship to the boundary threshold. Each of these patterns was replicated in a separate cohort of 92 subjects. For both drift rate and non-decision time (and in both the Discovery and Replication cohort), the relationship between cognitive function and integration was most pronounced across frontoparietal and subcortical regions (FDR $p < 0.05$; Figure 3b and 3c). Together, these results suggest that a globally efficient, integrated network architecture supports fast, effective computation throughout the cognitive processing stream (Krienen et al., 2014), potentially through the facilitation of parallel processing mechanisms (Sigman and Dehaene, 2008).

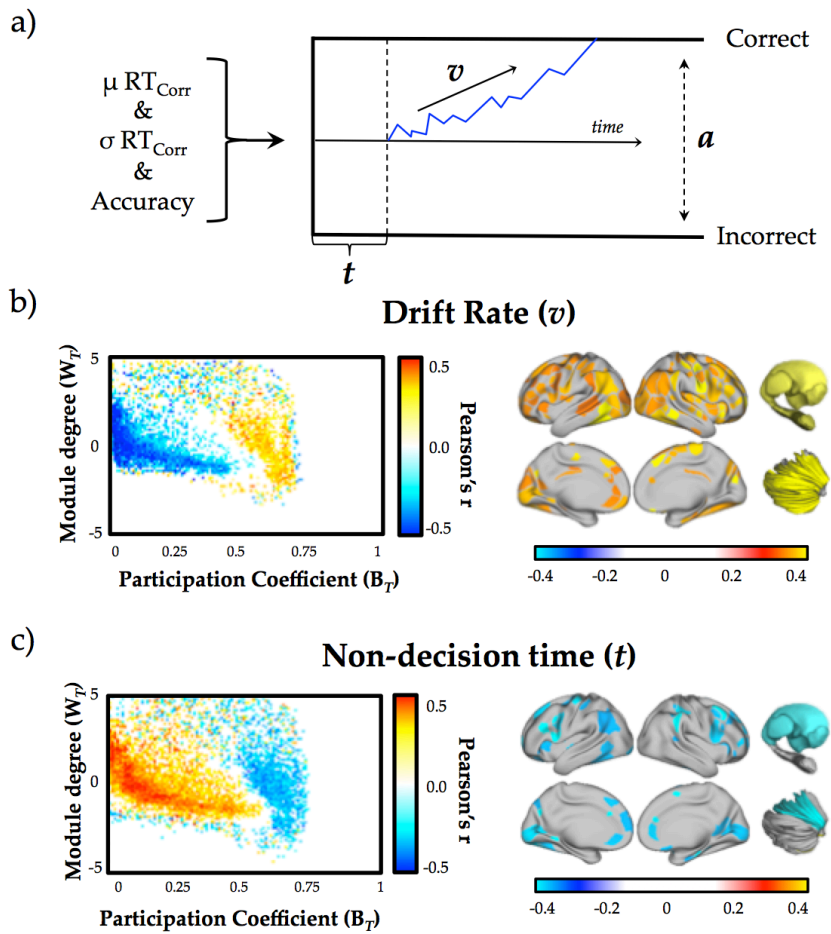

**Figure 3 – Relationship between cognitive acuity and the cartographic profile:** a) a graphical depiction of the drift-diffusion model, which uses the mean and standard deviation of a subjects reaction time and performance accuracy to estimate the 'drift rate', or rate of evidence accumulation ($v$), the length of non-decision time ($t$) and the response boundary ($a$); b) left – group-level correlation between drift rate on the N-back task and each bin of the mean cartographic profile during the N-back task in the Discovery cohort; right – parcels showing a positive correlation between mean $B_T$ and drift rate; and c) left – group-level correlation between non-decision time on the N-back task and each bin of the mean cartographic profile during the N-back task in the Discovery cohort; right – parcels showing a negative correlation between mean $B_T$ and non-decision time. False discovery rate, alpha = 0.05. No bins of the cartographic profile showed a consistent response with the response boundary. Similarly, no parcels showed a significant correlation between $W_T$ and any of the three diffusion model fits.

## Network Cartography Fluctuates with Pupil Diameter

Based on the results of these experiments, we hypothesized that neuromodulatory brain systems that mediate neural gain control (Aston-Jones and Cohen, 2005) may play an important role in regulating global integration. Recent invasive electrophysiological recordings in non-human primates have shown that pupil diameter tracks with neural firing in ascending neuromodulatory systems, such as the locus coeruleus, and as such, can be used as a surrogate measure of arousal and task engagement (McGinley et al., 2015). Therefore, we measured pupil diameter from individuals in a separate resting state dataset (14 individuals; TR = 2s; 3.5mm$^3$ voxels; 204 volumes; Murphy et al., 2014) and compared alterations in pupil diameter with the cartographic profile ($w$ = 10 TRs). As predicted, we observed a positive correlation between pupil diameter and mean $B_T$ (group mean r = 0.241 +/- 0.06; $R^2$ = 0.06; p = 10$^{-5}$; Figure 4), suggesting that the observed global fluctuations in network structure over time were significantly related to ongoing dynamic alterations in ascending neuromodulatory input to the cortex and subcortex.

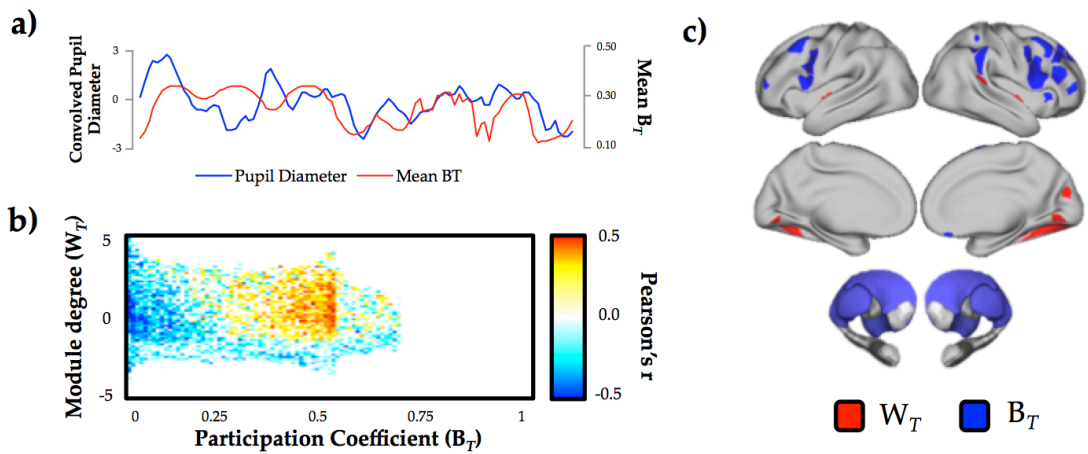

**Figure 4 – Relationship between cartography and pupillometry**: a) an example time series (subject #1) showing the covariance between the pupil diameter (after convolution with a hemodynamic response function; blue) and mean between-module connectivity ($B_T$; red); b) mean Pearson correlation between each bin of the cartographic profile and the convolved pupil diameter. Across the cohort of 14 subjects, we observed a positive relationship between pupil diameter and network-level integration (FDR p = 0.05); c) results from a conjunction analysis (FDR p < 0.05) that compared relationships between $W_T$ (red) or $B_T$ (blue) and drift-rate (positive correlation), non-decision time (inverse correlation) and pupillometry (positive correlation). There were no cerebellar parcels above threshold in all three contrasts.

**Identifying Regions Related to Global Integration**

To further investigate the neurobiological mechanisms responsible for fluctuations in network topology over time, we used a parcel-wise conjunction analysis (Nichols et al., 2005) to identify a set of regions that were significantly related to drift rate, non-decision time and pupil diameter. This analysis revealed a right-lateralized network of frontal, parietal, thalamic and striatal regions that were associated with consistently elevated $B_T$ across the three comparisons (blue; Figure 4c) and a set of regions in visual cortex and insula that were associated with elevated $W_T$ (red; Figure 4c). Together, these results highlight a distributed network of brain regions that mediate the computational integration required for effective cognitive processing.

**Reproducibility**

To test the reproducibility of our results, we performed three separate replication analyses: i) on a second resting state session from the same cohort of 92 unrelated subjects; ii) on a different cohort of 92 unrelated subjects from the HCP consortium; and iii) on 152 subjects from a separate dataset acquired at a different scanning site, using high-resolution functional data from the NKI Rockland dataset (Nooner et al., 2012). For each analysis in the resting state, we replicated the results described above on each subject and then summarized each outcome measure of interest at the group level (minimum r = 0.564; all p < 0.001; see *Materials and Methods*). In the task data, each of the relationships identified between the cartographic profile and behavior were replicated in the second set of 92 individuals from the HCP (Both r > 0.610; p < 0.001; Figure S3). These results suggest that the time-resolved measures identified in this study were reliable across sessions, individuals and independent datasets collected using different scanners and imaging protocols.

**Discussion**

In this manuscript, we mapped the spatiotemporal dynamics of complex network structure in the human brain, revealing a dynamical system that fluctuates between segregated and integrated network topology (Figure 1). The cartographic profile observed in the resting state was modulated by the performance of a range of cognitive tasks in proportion to task demands (Figure 2). Importantly, the extent to which the brain was integrated during the N-back task was correlated with faster drift rate and shorter non-decision time, suggesting that integration relates to fast and effective cognitive performance (Figure 3). We then showed that integration within the functional connectome correlated with increases in pupil diameter (Figure 4), highlighting a potential neurobiological mechanism responsible for modulating network-level dynamics in the human brain. Finally, we were able to demonstrate that a network of right-lateralized frontoparietal and subcortical regions were responsible for mediating the effects of integration on cognitive function (Figure 4c).

In our final experiment, we demonstrated that the fluctuations in network cartography in the resting state correlate with changes in pupil diameter (Figure 4), which itself is a marker of arousal and behavioral engagement (McGinley et al., 2015). The locus coeruleus (Aston-Jones and Cohen, 2005) is known to modulate pupil diameter (Joshi et al., 2016), and thus by inference, may play a role in the modulation of fluctuations in global network topology. Thus, our results extend previous studies that have demonstrated a crucial link between neural gain and functional connectivity (Eldar et al., 2013; Yellin et al., 2015) by showing that fluctuations in neural gain are linked to alterations in network topology that directly relate to effective behavioral performance. Indeed, there is a wealth of evidence to suggest that neuromodulatory inputs can have complex, non-linear effects on network organization and behavior (Bargmann and Marder, 2013), perhaps as a result of the balance between the 'top-down' attentional modulation of network architecture (Sara, 2009) and 'bottom-up' neuromodulatory input from the brainstem (Safaai et al., 2015). The network of right-lateralized cortical regions consistently associated with elevations in integration in our study provides further support for this hypothesis (Figure 4c), as ascending noradrenergic inputs preferentially impact neural function within the right cortical hemisphere (Pearlson and Robinson, 1981). Importantly, the topological organization of the functional connectome is likely to arise as the end result of multiple competing factors, including changes in tone within other neuromodulatory systems, such as the basal cholinergic nuclei (Steriade and McCarley, 2013), local interactions among functional regions, and activity in other diffuse projection systems, such as the intralaminar thalamic nuclei (Van der Werf et al., 2002).

Irrespective of the precise mechanism driving global fluctuations, our results suggest that system-wide alterations in network topology facilitate more effective behavioral performance, a hypothesis that has already garnered support from studies both in network dynamics (Kitzbichler et al., 2011) and pupillometery (Murphy et al., 2016). There is now growing evidence to support the notion that the brain traverses a metastable state-space in time (Deco et al., 2015), balancing the opposing tendencies for specialized, segregated processing with the need for global coordination and integration (Tognoli and Kelso, 2014). Here, we extend these conceptual studies by demonstrating fluctuations in network topology consistent with the concept of metastability that relate to effective behavioral performance.

Although we were able to demonstrate that greater system-wide integration was associated with improved cognitive performance on an N-back task, the precise role of network topology in cognition requires further exploration. While the N-back task is often used as an index of general cognitive function, cognition is a complex construct that is comprised of dissociable sub-components, such as updating, set-shifting and response inhibition (Miyake et al., 2000), that likely rely on overlapping, yet distinct, neural architectures (Duncan, 2010; Poldrack et al., 2011). The relationship between global integration and each of these distinct cognitive sub-components remains an open question for future studies. Furthermore, it will be important to determine whether fluctuations in the cartographic profile are sensitive to trial-by-trial performance on specific cognitive tasks, as others have shown with different measures of functional connectivity and network topology (Gonzalez-Castillo et al., 2015; Sadaghiani et al., 2015). It will also be crucial to clarify the relationship between fluctuations in network structure and the dynamic interactions between endogenous and exogenous attentional systems (Fox et al., 2005; Mason et al., 2007), which are presumed to be related to dynamics involving key hubs within the default network (Vatansever et al., 2015), such as the precuneus (Leech et al., 2012; Lin et al., 2016; Utevsky et al., 2014).

There are also some important limitations to note in our study. Firstly, although we provide indirect evidence for the relationship between neural gain and effective cognitive performance, the direct relationship between ascending neuromodulatory input to the brain and network topology requires further confirmation, perhaps utilizing the temporal resolution afforded by electrophysiological measures of brain function. In addition, although we attempted to directly compare the MTD approach to the standard approach used to calculate time-resolved connectivity (namely, sliding window Pearson's correlation), it bears mention that there are many techniques used to estimate these measures (Hutchison et al., 2012) and as such, further work is required to determine the robustness of the fluctuations in network topology across multiple time-sensitive connectivity metrics.

Together, our results demonstrate that global brain integration is closely related to cognitive function during an N-back task. By catalyzing communication between specialist regions of the brain that would otherwise remain segregated, global integration increases an individuals ability to accomplish complex cognitive tasks, potentially accelerating behavioral innovation and improving fitness in novel scenarios (Shanahan, 2012). As such, global integration is an important candidate mechanism responsible for the evolution of complex brain networks (van den Heuvel et al., 2016), and hence, for explaining the mechanism through which the brain creates complex, adaptive behavior.

**Materials and Methods**

**Data acquisition**

For the primary discovery analysis, minimally preprocessed resting fMRI data were acquired from 100 unrelated participants from the Human Connectome Project (mean age 29.5 years, 55% female) (Glasser et al., 2013). For each participant, 14 minutes 30 seconds of resting state data were acquired using multiband gradient-echo echoplanar imaging. The following parameters were used for data acquisition: relaxation time (TR) = 720 ms, echo time = 33.1 ms, multiband factor = 8, flip angle = 52 degrees, field of view = 208x180 mm (matrix = 104 x 90), 2x2x2 isotropic voxels with 72 slices, alternated LR/RL phase encoding.

In addition to the discovery analysis, we also performed an extensive series of replication analyses including: i) data from the same participants using resting state data acquired during a second rest scan during the same scanning session; ii) an independent cohort of 100 unrelated participants from the HCP dataset using identical acquisition parameters at the same scanning site; and iii) an out-of-sample replication using data collected from the NKI Rockland sample (TR = 650msec; voxel-size 3mm$^3$) as part of the 1000 Functional Connectomes Project (Nooner et al., 2012).

**Data pre-processing**

Bias field correction and motion correction (12 linear DOF using FSL's FLIRT) were applied to the HCP resting state data as part of the minimal preprocessing pipeline (Glasser et al., 2013). The first 100 time points were discarded from the data due to the presence of an evoked auditory signal associated with noise in the scanner. Resting state data acquired from the NKI Rockland sample were realigned to correct for head motion and then each participants' functional scans were registered to both their T1-weighted structural image and then to the MNI152 atlas using boundary based registration (http://fsl.fmrib.ox.ac.uk/fsl/fslwiki/) and Advanced Normalization Tools software (Avants et al., 2008). After co-registration, data were manually inspected and of the 173 original participants, 11 [6.3%] scans were discarded due insufficient coverage of orbitofrontal cortex, temporopolar cortex and/or cerebellum.

Temporal artifacts were identified in each dataset by calculating framewise displacement (FD) from the derivatives of the six rigid-body realignment parameters estimated during standard volume realignment(Power et al., 2014), as well as the root mean square change in BOLD signal from volume to volume (DVARS). Frames associated with FD > 0.5mm or DVARS > 5% were identified, and participants with greater than 20% of the resting time points exceeding these values were excluded from further analysis (HCP group 1: 8/100; HCP group 2: 8/100; NKI group: 10/162). Due to concerns associated with the alteration of the temporal structure of the images, the data used in the main analysis were not 'scrubbed' (Power et al., 2014), however we did explicitly compare the results of our experiment with scrubbed data (missing values were corrected using interpolation) and found strong correspondence between the outcome measures of the two studies (see *Validation*). Following artifact detection, nuisance covariates associated with the 12 linear head movement parameters (and their temporal derivatives), FD, DVARS, and anatomical masks from the cerebral spinal fluid and deep cerebral white matter were regressed from the data using the CompCor strategy (Behzadi et al., 2007). Finally, in keeping with previous time-resolved connectivity experiments (Bassett et al., 2015), a temporal band pass filter (0.071 < f < 0.125 Hz) was applied to the data (see *Validation*).

**Brain parcellation**

Following pre-processing, the mean time series was extracted from 375 pre-defined regions-of-interest (ROI). To ensure whole-brain coverage, we extracted: 333 cortical parcels (161 and 162 regions from the left and right hemispheres, respectively) using the Gordon atlas (Gordon et al., 2014), 14 subcortical regions from Harvard-Oxford subcortical atlas (bilateral thalamus, caudate, putamen, ventral striatum, globus pallidus, amygdala and hippocampus; http://fsl.fmrib.ox.ac.uk/), and 28 cerebellar regions from the SUIT atlas (Diedrichsen et al., 2009) for each participant in the study. These ROIs were chosen to maximize our ability to interrogate fluctuations in network architecture over time, however it bears mention that the regions do not necessarily reflect a 'ground truth' parcellation, as functional divisions may differ across subjects (Laumann et al., 2015) and subdivisions of the subcortical structures may show differential patterns of network involvement over time.

**Multiplication of temporal derivatives**

To estimate functional connectivity between the 375 ROIs, we used a recently described statistical technique (Multiplication of Temporal Derivatives; MTD – http://github.com/macshine/coupling/) (Shine et al., 2015) that allows greater temporal resolution of time-resolved connectivity in BOLD time series data when compared to the conventional sliding-window Pearson's correlation coefficient (Shine et al., 2015). The MTD is computed by calculating the point-wise product of temporal derivative of pairwise time series (Equation 1). In order to reduce the contamination of high-frequency noise in the time-resolved connectivity data, the

MTD is averaged by calculating the mean value over a temporal window, $w$ (https://github.com/macshine/coupling/).

$$MTD_{ijt} = \frac{1}{w}\sum_{t}^{t+w} \frac{(dt_{it} \times dt_{jt})}{(\sigma_{dt_i} \times \sigma_{dt_j})} \qquad [1]$$

**Equation 1** – Multiplication of Temporal Derivatives, where for each time point, $t$, the MTD for the pairwise interaction between region $i$ and $j$ is defined according to equation 1, where $dt$ is the first temporal derivative of the $i^{th}$ or $j^{th}$ time series at time $t$, $\sigma$ is the standard deviation of the temporal derivative time series for region $i$ or $j$ and $w$ is the window length of the simple moving average. This equation can then be calculated over the course of a time series to obtain an estimate of time-resolved connectivity between pairs of regions.

**Time-resolved functional connectivity**

Time-resolved functional connectivity was calculated between all 375 brain regions using the MTD (Shine et al., 2015) within a sliding temporal window of 14 time points (10.1 seconds for HCP; 16 time points for NKI data ~ 10.4 seconds). Individual functional connectivity matrices were calculated within each temporal window, thus generating an unthresholded (that is, signed and weighted) 3D adjacency matrix (region × region × time) for each participant. Previous work has shown that, when using the MTD, a window length of seven time points provides optimal sensitivity and specificity for detecting dynamic changes in functional connectivity structure in simulated time series data (Shine et al., 2015). To balance these benefits with the need to track changes in slow cortical fluctuations which are hypothesized to fluctuate at ~0.1 Hz (Shen et al., 2015), we used a temporal window of 14 time points to calculate a simple moving average of the MTD, which allowed for estimates of signals at approximately 0.1 Hz. While there are statistical arguments to suggest that the potential effects of noise can render estimation of connectivity matrices difficult with smaller samples, it is currently unclear whether these issues will have the same effects on the covariance estimates created with the MTD. However, we note that the MTD is more sensitive to changes in covariance than connectivity (Shine et al., 2015) and others have shown that covariance is a more reliable marker of coupling strength in BOLD data (Cole et al., 2016). Most importantly, as we show, our analyses were reliable and replicable using the MTD across multiple datasets.

**Time- resolved community structure**

The Louvain modularity algorithm from the Brain Connectivity Toolbox (BCT; (Rubinov and Sporns, 2010)) was used in combination with the MTD to estimate both time-averaged and time-resolved community structure. The Louvain algorithm iteratively maximizes the modularity statistic, Q, for different community assignments until the maximum possible score of Q has been obtained (see Equation 2). The modularity estimate for a given network is therefore a quantification of the extent to which the network may be subdivided into communities with stronger within-module than between-module connections.

$$Q_T = \frac{1}{v^+}\sum_{ij}(w_{ij}^+ - e_{ij}^+)\delta_{M_iM_j} - \frac{1}{v^+ + v^-}\sum_{ij}(w_{ij}^- - e_{ij}^-)\delta_{M_iM_j} \qquad [2]$$

**Equation 2** – Louvain modularity algorithm, where $v$ is the total weight of the network (sum of all negative and positive connections), $w_{ij}$ is the weighted and signed connection between regions $i$ and $j$, $e_{ij}$ is the strength of a connection divided by the total weight of the network, and $\delta_{M_iM_j}$ is set to 1 when regions are in the same community and 0 otherwise. '+' and '−' superscripts denote all positive and negative connections, respectively.

For each temporal window, the community assignment for each region was assessed 500 times and a consensus partition was identified using a fine-tuning algorithm from the Brain Connectivity Toolbox (BCT, http://www.brain-connectivity-toolbox.net/). This then afforded an estimate of both the time-resolved modularity ($Q_T$) and cluster assignment ($Ci_T$) within each temporal window for each participant in the study. All graph theoretical measures were calculated on weighted and signed connectivity matrices, such that no arbitrary thresholding was required (Rubinov and Sporns, 2010).

Based on time-resolved community assignments, we estimated within-module connectivity by calculating the time-resolved module-degree Z-score ($W_T$; within module strength) for each region in our analysis (Equation 3(Guimerà and Nunes Amaral, 2005).

$$W_{iT} = \frac{\kappa_{iT} - \acute{\kappa}_{s_{iT}}}{\sigma_{\kappa_{s_{iT}}}} \qquad [3]$$

**Equation 3** – Module degree Z-score, $W_{iT}$, where $\kappa_{iT}$ is the strength of the connections of region $i$ to other regions in its module $s_i$ at time $T$, $\acute{\kappa}_{s_{iT}}$ is the average of $\kappa$ over all the regions in $s_i$ at time $T$, and $\sigma_{\kappa_{s_{iT}}}$ is the standard deviation of $\kappa$ in $s_i$ at time $T$.

**Time- resolved hub structure**

The participation coefficient, $B_T$, quantifies the extent to which a region connects across all modules (i.e. between-module strength) and has previously been used to characterize hubs within brain networks (e.g. see (Power et al., 2013). The $B_T$ for each region was calculated within each temporal window using Equation 4.

$$B_{iT} = 1 - \sum_{s=1}^{n_M} \left(\frac{\kappa_{isT}}{\kappa_{iT}}\right)^2 \qquad [4]$$

**Equation 4** - Participation coefficient $B_{iT}$, where $\kappa_{isT}$ is the strength of the positive connections of region $i$ to regions in module $s$ at time $T$, and $\kappa_{iT}$ is the sum of strengths of all positive connections of region $i$ at time $T$. The participation coefficient of a region is therefore close to 1 if its connections are uniformly distributed among all the modules and 0 if all of its links are within its own module.

**Cartographic profiling**

To track fluctuations in cartography over time, we created a novel analysis technique that did not require the labeling of each node into a pre-defined cartographic class (Guimerà and Nunes Amaral, 2005). For each temporal window, we computed a joint histogram of within- and between-module connectivity measures, which we refer to here as a "cartographic profile" (Figure 1). Code for this analysis is freely available at https://github.com/macshine/integration/. To test whether the cartographic profile of the resting brain fluctuated over time between two topological extremes, we performed clustering of temporal windows without the use of cartographic class labels. To do so, we classified the joint histogram of each temporal window (which is naïve to cartographic boundaries) over time using a k-means clustering analysis ($k$ = 2). As a result of this analysis, each window was assigned to one of two clusters. K-means was repeated with 500 random restarts (i.e. replicates as implemented in the MATLAB *k-means* function) to mitigate the sensitivity of k-means to initial conditions. To ensure that the *a priori* choice of two clusters for the k-means analysis was reflective of the broader patterns in the data across multiple values of $k$, we re-ran the clustering analysis in the discovery cohort of 92 subjects across a range of $k$ (2-20) and then compared the resultant cluster partitions to the $k$ = 2 clusters by calculating the mutual information between the each pair of partitions. The partition identified at each value of *k* was strongly similar to the pattern identified at *k* = 2 (mean mutual information = 0.400 ± 0.02; Figure S1). We also provided further evidence for this partition by performing a principle component analysis for each subject's data – this test demonstrated that the first two principle component for each subject were associated with the integrated (20.2 ± 1.4% variance) or segregated state (4.9 ± 2.3% of variance). Together, these results suggest that our choice of utilizing k = 2 to cluster the joint histograms over time was reflective of a relatively "natural" clustering pattern in the data.

To explicitly test whether the resting brain fluctuated more frequently than a stationary null model, we calculated the absolute value of the window-to-window difference in the mean $B_T$ score for each iteration of the VAR null model. In keeping with Zalesky et al. (Zalesky et al., 2014), VAR model order was set at 11, appropriately mimicking the expected temporal signature of the BOLD response in 0.72s TR data. The mean covariance matrix across all 92 subjects from the discovery group was used to generate 2500 independent null data sets, which allows for the appropriate estimation of the tails of non-parametric distributions (Nichols and Holmes, 2002). These time series were then filtered in a similar fashion to the BOLD data. For each analysis, the maximum statistic was concatenated for each independent simulation. We then calculated the 95[th] percentile of this distribution and used this value to determine whether the resting state data fluctuated more frequently than the null model. In the discovery cohort, 16.1 ± 1.1% of temporal windows were associated with deviations ≥ 95[th] percentile of the VAR null model (i.e. greater than the predicted 5%), suggesting that the resting state was associated with significant dynamic fluctuations in topology. Importantly, the significant fluctuations along the $B_T$ axis remained after correcting for ongoing changes in the number of modules per temporal window.

To estimate patterns of topology associated with each state, the original 3D connectivity matrix containing MTD values was then reorganized into those windows associated with the two states. The modularity of each window was then calculated using the Louvain algorithm (Equation 2) and the resultant values were then compared statistically using an independent samples t-test. Importantly, the two states were matched on graph density, suggesting that the fluctuations in $B_T$ did not occur simply due to alterations in network sparsity over time. A similar technique was used to estimate the global efficiency of each temporal window. As global efficiency (Equation 5) cannot be computed from networks with negative weights (Barch et al., 2013), we first thresholded the connectivity matrix within each window to include only positive edge weights before calculating global efficiency. The estimate of modularity and efficiency within each temporal window was then compared between those windows associated with the integrated and segregated states using an independent samples t-test (Figure S2).

$$E_{glob} = \frac{2}{n(n=1)} \sum_{i<j \in G}^{n} \frac{1}{d_{i,j}} \qquad [5]$$

**Equation 4** – global efficiency of a network, where *n* denotes the total nodes in the network and $d_{i,j}$ denotes the shortest path between a node *i* and neighboring node *j*.

To estimate the patterns of brain connectivity associated with each state, we binned each region's $W_T$ and $B_T$ scores into those windows associated with either the Integrated or Segregated state (using the *k* = 2 partition). We then compared the regional $W_T$ and $B_T$ scores across the two states using an independent-samples t-test. As expected, all 375 parcels demonstrated higher $B_T$ in the more Integrated state, whereas none of the 375 parcels showed significantly different $W_T$ in either state (FDR p < 0.05). For interpretation and display, regional BT scores were converted into Z-scores and then projected onto surface renderings (Figure 1).

**Task-based alterations in the cartographic profile**

To assess task-based functional connectivity, preprocessed data from the original 92 unrelated subjects from the discovery cohort were collected while these subjects performed seven different tasks in the fMRI (Barch et al., 2013): i) a simple motor task in which the participants were presented with visual cues that required them to tap their left or right fingers, squeeze their left or right toes, or move their tongue; ii) a visually-based N-back task, which consisted of interleaved 10 second blocks of a high (2-back) and low (0-back) load N-back task, each block with object stimuli in one of four classes (places, faces, body parts and tools; iii) a social cognition task, in which subjects passively viewed videos of interacting objects and were asked to judge the character of their interactions; iv) a gambling task, which took the form of a card 'guessing' game in which subjects were rewarded for correct responses; v) a relational matching task, in which subjects were required to distinguish between items that were either related to one another conceptually or had a matching pattern; vi) a emotional processing task, in which subjects are asked to judge the emotional character of faces; and vi) a language task, in which subjects either listened to short narratives or performed a simple mathematical task. See (Barch et al., 2013) for further details of each experimental paradigm.

The mean time series was then extracted from the same 375 regions as defined in the resting state analysis. To control for spurious patterns of connectivity associated with task-evoked activity, we first regressed the HRF-convolved task block data from each time series. The MTD metric was then calculated on the residuals of this regression using a window length of 14 TRs (~10 seconds at 0.72 second TR). These data were then subjected to a cartographic profiling analysis in a similar fashion to the resting state data.

To compare the patterns of time-resolved connectivity across the N-back task to those observed during rest, we tested whether any bins within the 2-dimensional cartographic profile were significantly modulated by task by running a mixed-effects General Linear Model analysis at the individual level, fitting the group-averaged joint histogram to regressors tracking two-back, zero-back and rest blocks in both the Motor and the N-back task, separately. We then compared the task blocks and the resting state data statistically using separate two-sided, one-sample t-tests across subjects (FDR $p < 0.05$). We observed a rightward deviation in the mean cartographic profile during the 2-back vs 0-back block, however to allow direct comparison across tasks and rest, we opted to include the mean 2-back profile for each comparison described in the main manuscript. A similar analysis was run comparing the mean $W_T$ and $B_T$ across all 375 parcels. As in previous steps, the regional $B_T$ scores were converted into Z-scores (otherwise the regional heterogeneity associated with each task would be hidden within the much-larger mean effect) and then projected onto surface renderings (Figure 2).

To ensure that any differences observed during task performance were not confounded by fluctuations in global signal or connectivity, we replicated the analysis after separately regressing the global signal and the mean MTD value across all parcels. The relationship between the task regressor and the mean BT survived in each case (global signal: mean r = 0.452 ± 0.21; mean MTD: mean r = 0.393 ± 0.14; p = $10^{-5}$).

In order to assess the alteration in the cartographic profile as a function of task performance, we estimated the affine transformation (using a correlation cost function with 3 degrees of freedom, including translation and rotation parameters) between each individual subjects' resting state cartographic profile and the profile observed in each of the seven tasks.

**Investigating the Relationship Between Cartography and Behavior**

To interrogate the relationship between the cartographic profile and behavioral performance, we fit an EZ-diffusion model to the performance measures from the N-back task (Wagenmakers et al., 2007). This model takes in the mean RT on correct trials, mean variance of RT across correct trials, and mean accuracy across the task and computes from them a value for drift rate, boundary separation, and non-decision time – the three main parameters for the diffusion model (Figure 3). We used the EZ-diffusion model instead of alternative diffusion fitting routines (e.g. fast-dm or DMAT) because previous work has shown that the EZ-diffusion model is particularly effective for recovering individual differences in parameter values, as we're interested in here (van Ravenzwaaij and Oberauer, 2009). After fitting each subjects data to the diffusion model, we then performed a group-level Pearson's correlation between each bin of the mean joint histogram in each task and the three outcome measures associated with the N-back task: the drift rate (Figure 3b), the non-decision time (Figure 3c) and the boundary threshold (results not shown, as no bins survived multiple comparisons correction). The model was fit on results from the 2-back task blocks, as a many subjects made no errors on the 0-back condition, thus precluding our ability to fit their data to the parameters of the drift diffusion model. For each comparison, the null hypothesis of no relationship was rejected after false discovery rate correction ($p < 0.05$). We also compared the cartographic profile with median reaction time and accuracy for both the cohorts and observed a similar relationship between integration and improved performance.

Some work suggests that the EZ-diffusion model performs poorly when there are "contaminants" in the data (Ratcliff et al., 2015), which are trials in which the usual diffusion parameters do not apply (like fast guesses and attentional lapses). We searched for evidence of contaminants in our data and found no evidence of them (i.e. few responses were extremely short [<300 msec] and the few that existed were statistically above chance). Therefore, we proceeded with the EZ-diffusion model, which performs as well or better than more complicated fitting routines when contaminants are not present (Ratcliff et al., 2015; van Ravenzwaaij and Oberauer, 2009).

**Network Cartography Fluctuates with Pupil Diameter**

To test the hypothesis that fluctuations in cartography related to activity in ascending neuromodulatory systems, we acquired a separate dataset of 14 individuals (mean age: 29 years; 8/14 male) in which pupil diameter was measured over time during the quiet resting state (TR = 2s; 3.5mm$^3$ voxels; 204 volumes) (Murphy et al., 2014). Participants were instructed to relax, think of nothing in particular and maintain fixation for 8 min at a centrally presented crosshair (subtending $0.65^0$ of the visual angle). BOLD fMRI data were preprocessed using SPM8 software (www.fil.ion.ucl.ac.uk/spm). Pupil diameter was recorded continuously from the left eye at rest and during task using an

iView X MRI-SV eyetracker (SMI, Needham, MA) at a sampling rate of 60 Hz. Pupillometric data were thoroughly pre-processed to remove potential sources of noise (see (Murphy et al., 2014) for details) and then down-sampled to a 0.5 Hz sampling rate (in order to match the sampling frequency of the fMRI data). A pupil diameter vector for each scanning run was then convolved with the informed basis set to yield three pupil regressors of interest per participant. The mean of these regressors was then correlated with the cartographic profile across all temporal windows for each of the 14 subjects (mean correlation: r = 0.241 ± 0.06). A set of one-sample t-tests was then used to test whether the correlation between each bin of the cartographic profile was significantly different from zero (FDR $p < 0.05$). A similar t-test was used to determine whether the correlation between the mean $B_T$ and pupil diameter was significantly greater than zero across the cohort of 14 subjects.

**Identifying Regions Related to Global Integration**

We used a parcel-wise conjunction analysis (Nichols et al., 2005) to identify a set of regions in which the $B_T$ and $W_T$ were significantly related to drift rate, non-decision time and pupil diameter. For each comparison in turn, we determined whether the $W_T/B_T$ individual parcel was significantly correlated with each outcome measure of interest above chance (FDR $p < 0.05$). We then binarized the resultant parcel vectors and calculate a conjunction analysis, separately for both $W_T$ and $B_T$. Results were then projected onto surface renderings for interpretation.

**Replication analysis**

To quantify how well our results replicated across sessions and datasets, we calculated group-level correlations between each of the measures identified in our analysis. Overall, we observed a strong positive correlation between the outcome measures identified in the two sessions (for all statistical tests, p < 0.001): graph measures – $r_{W_T}$ = 0.982, $r_{B_T}$ = 0.957; and mean cartographic profiles $r_{Cart}$ = 0.982 (Figure S3). We also confirmed the presence of these results in a unique cohort of 92 unrelated participants from the HCP: graph measures – $r_{W_T}$ = 0.971, $r_{B_T}$ = 0.967; and mean cartographic profiles – $r_{Cart}$ = 0.973 (Figure S3). We also observed similarly positive relationships between the group-level outcome measures estimated from the HCP and NKI data (for all statistical tests, p < 0.001): graph measures – $r_{W_T}$= 0.941, $r_{B_T}$ = 0.857; and mean cartographic profiles – $r_{Cart}$ = 0.927 (Figure S3). In addition, the same fluctuations observed in the HCP dataset were also present in the NKI dataset (see Video 3 at http://github.com/macshine/coupling).

Finally, the linear relationships between behavioral performance and the cartographic profile were consistent across the discovery and replication datasets. A spatial correlation between the two datasets was strongly positive for both the relationship with drift rate (r = 0.613; $R^2$ = 0.37; p = $10^{-11}$; Figure S4) and non-decision time (r = 0.681; $R^2$ = 0.46; p = $10^{-15}$; Figure S4), but the null hypothesis could not be rejected for the diffusion boundary (p > 0.500).

**Acknowledgements**

The data reported in this paper were made publicly available by the Human Connectome Project and 1000 Functional Connectomes project. There were no conflicts of interest for any authors. JMS, PGB, OK, PTB and RAP designed the analysis. JMS performed the analysis and wrote the first draft. JMS, PTB, OK, PGB, JHB, KJG, CAM and RAP reviewed and edited the manuscript. We would like to thank Vanessa Sochat for her assistance with implementation, Peter Murphy, Redmond O'Connell and Ian Robertson for their work collecting and analyzing the data used for the pupillometry analysis, Jamie Li for help with data analysis, Michael Riis and Ian Eisenberg for their helpful insights and Tim Laumann, Claire O'Callaghan, Rick Shine and Rav Suri for their critical review of the manuscript.

# References

Aston-Jones, G., Cohen, J.D., 2005. An Integrative Theory of Locus Coeruleus-Norepinephrine Function: Adaptive Gain and Optimal Performance. Annu. Rev. Neurosci. 28, 403–450.

Avants, B.B., Epstein, C.L., Grossman, M., Gee, J.C., 2008. Symmetric diffeomorphic image registration with cross-correlation: evaluating automated labeling of elderly and neurodegenerative brain. Med Image Anal 12, 26–41.

Barch, D.M., et. al., 2013. Function in the human connectome: task-fMRI and individual differences in behavior. NeuroImage 80, 169–189.

Bargmann, C.I., Marder, E., 2013. From the connectome to brain function. Nat Meth.

Barttfeld, P., et. al., 2015. Signature of consciousness in the dynamics of resting-state brain activity. Proc. Natl. Acad. Sci. U.S.A. 112, 887–892.

Bassett, D.S., et. al., 2011. Dynamic reconfiguration of human brain networks during learning. Proc. Natl. Acad. Sci. U.S.A. 108, 7641–7646.

Bassett, D.S., Yang, M., Wymbs, N.F., Grafton, S.T., 2015. Learning-induced autonomy of sensorimotor systems. Nat Neurosci 18, 744–751.

Behzadi, Y., Restom, K., Liau, J., Liu, T.T., 2007. A component based noise correction method (CompCor) for BOLD and perfusion based fMRI. NeuroImage 37, 90–101.

Betzel, R.F., Fukushima, M., He, Y., Zuo, X.-N., Sporns, O., 2015. Dynamic fluctuations coincide with periods of high and low modularity in resting-state functional brain networks. NeuroImage 127, 287–297.

Bullmore, E., Sporns, O., 2012. The economy of brain network organization. Nat. Rev. Neurosci. 13, 336–349.

Cole, M.W., Bassett, D.S., Power, J.D., Braver, T.S., Petersen, S.E., 2014. Intrinsic and task-evoked network architectures of the human brain. Neuron 83, 238–251.

Cole, M.W., Yang, G.J., Murray, J.D., Repovs, G., 2016. Functional connectivity change as shared signal dynamics. Journal of neuroscience methods. 259:22-39.

Corbetta, M., Shulman, G.L., 2002. Control of Goal-directed and stimulus-driven attention in the brain. Nat. Rev. Neurosci. 3, 215–229.

Deco, G., Tononi, G., Boly, M., Kringelbach, M.L., 2015. Rethinking segregation and integration: contributions of whole-brain modelling. Nat. Rev. Neurosci. 16, 430–439.

Diedrichsen, J., Balsters, J.H., Flavell, J., Cussans, E., Ramnani, N., 2009. A probabilistic MR atlas of the human cerebellum. NeuroImage 46, 39–46.

Duncan, J., 2010. The multiple-demand (MD) system of the primate brain: mental programs for intelligent behaviour. Trends in Cognitive Sciences 14, 172–179.

Eldar, E., Cohen, J.D., Niv, Y., 2013. The effects of neural gain on attention and learning. Nat Neurosci. 16:1146-53.

Fox, M.D., et. al, 2005. From The Cover: The human brain is intrinsically organized into dynamic, anticorrelated functional networks. Proceedings of the National Academy of Sciences 102, 9673–9678.

Fries, P., 2015. Rhythms for Cognition: Communication through Coherence. Neuron 88, 220–235.

Glasser, M.F., et. al., 2013. The minimal preprocessing pipelines for the Human Connectome Project. NeuroImage 80, 105–124.

Gonzalez-Castillo, J., et. al, 2015. Tracking ongoing cognition in individuals using brief, whole-brain functional connectivity patterns. Proc. Natl. Acad. Sci. U.S.A. 112, 8762–8767.

Gordon, E.M., et. al., 2014. Generation and Evaluation of a Cortical Area Parcellation from Resting-State Correlations. Cereb. Cortex. 26:288-303.

Guimerà, R., Nunes Amaral, L.A., 2005. Functional cartography of complex metabolic networks. Nature 433, 895–900.

Hutchison, R.M., Womelsdorf, T., Gati, J.S., Everling, S., Menon, R.S., 2012. Resting-state networks show dynamic functional connectivity in awake humans and anesthetized macaques. Hum Brain Mapp 34, 2154–2177.

Joshi, S., Li, Y., Kalwani, R.M., Gold, J.I., 2016. Relationships between Pupil Diameter and Neuronal Activity in the Locus Coeruleus, Colliculi, and Cingulate Cortex. Neuron 89, 221–234.

Kitzbichler, M.G., Henson, R.N.A., Smith, M.L., Nathan, P.J., Bullmore, E.T., 2011. Cognitive effort drives workspace configuration of human brain functional networks. J. Neurosci. 31, 8259–8270.

Krienen, F.M., Yeo, B.T.T., Buckner, R.L., 2014. Reconfigurable task-dependent functional coupling modes cluster around a core functional architecture. Philos. Trans. R. Soc. Lond., B, Biol. Sci. 369, 20130526–20130526.

Laumann, T.O., Gordon, E.M., Adeyemo, B., Snyder, A.Z., 2015. Functional System and Areal Organization of a Highly Sampled Individual Human Brain. Neuron. 87:657-70.

Leech, R., Braga, R., Sharp, D.J., 2012. Echoes of the brain within the posterior cingulate cortex. J. Neurosci. 32, 215–222.

Lin, P., et. al., 2016. Static and dynamic posterior cingulate cortex nodal topology of default mode network predicts attention task performance. Brain Imaging and Behavior 10, 212–225.

Mason, M.F., Norton, M.I., Van Horn, J.D., Wegner, D.M., Grafton, S.T., Macrae, C.N., 2007. Wandering Minds: The Default Network and Stimulus-Independent Thought. Science 315, 393–395.

Mattar, M.G., Cole, M.W., Thompson-Schill, S.L., Bassett, D.S., 2015. A Functional Cartography of Cognitive Systems. PLoS Comput. Biol. 11, e1004533.

McGinley, M.J., Vinck, M., Reimer, J., Batista-Brito, R., 2015. Waking state: rapid variations modulate neural and behavioral responses. Neuron.

Miyake, A., et. al., 2000. The Unity and Diversity of Executive Functions and Their Contributions to Complex "Frontal Lobe" Tasks: A Latent Variable Analysis. Cognitive Psychology 41, 49–100.

Murphy, P.R., O'Connell, R.G., O'Sullivan, M., Robertson, I.H., Balsters, J.H., 2014. Pupil diameter covaries with BOLD activity in human locus coeruleus. Hum Brain Mapp 35, 4140–4154.

Murphy, P.R., van Moort, M.L., Nieuwenhuis, S., 2016. The Pupillary Orienting Response Predicts Adaptive Behavioral Adjustment after Errors. PLoS ONE 11, e0151763.

Nichols, T., Brett, M., Andersson, J., Wager, T., Poline, J.-B., 2005. Valid conjunction inference with the minimum statistic. NeuroImage 25, 653–660.

Nichols, T.E., Holmes, A.P., 2002. Nonparametric permutation tests for functional neuroimaging: a primer with examples. Hum Brain Mapp 15, 1–25.

Nooner, K.B., et. al., 2012. The NKI-Rockland Sample: A Model for Accelerating the Pace of Discovery Science in Psychiatry. Front. Neurosci. 6, 152.

Pearlson, G.D., Robinson, R.G., 1981. Suction lesions of the frontal cerebral cortex in the rat induce asymmetrical behavioral and catecholaminergic responses. Brain Research 218, 233–242.

Poldrack, R.A., et al., 2011. The Cognitive Atlas: Toward a Knowledge Foundation for Cognitive Neuroscience. Front Neuroinform 5, 17.

Power, J.D., Mitra, A., Laumann, T.O., Snyder, A.Z., Schlaggar, B.L., Petersen, S.E., 2014. Methods to detect, characterize, and remove motion artifact in resting state fMRI. NeuroImage 84, 320–341.

Power, J.D., et. al., 2013. Evidence for hubs in human functional brain networks. Neuron 79, 798–813.

Ratcliff, R., 1978. A theory of memory retrieval. Psychological review. 85:59-108.

Ratcliff, R., Thompson, C.A., McKoon, G., 2015. Modeling individual differences in response time and accuracy in numeracy. Cognition. 137:115-36.

Rubinov, M., Sporns, O., 2010. Complex network measures of brain connectivity: Uses and interpretations. NeuroImage 52, 1059–1069.

Sadaghiani, S., Poline, J.-B., Kleinschmidt, A., D'Esposito, M., 2015. Ongoing dynamics in large-scale functional connectivity predict perception. Proc. Natl. Acad. Sci. U.S.A. 112, 8463–8468.

Safaai, H., Neves, R., Eschenko, O., Logothetis, N.K., Panzeri, S., 2015. Modeling

the effect of locus coeruleus firing on cortical state dynamics and single-trial sensory processing. Proc. Natl. Acad. Sci. U.S.A. 112, 12834–12839.

Sara, S.J., 2009. The locus coeruleus and noradrenergic modulation of cognition. Nat. Rev. Neurosci. 10, 211–223.

Shanahan, M., 2012. The brain's connective core and its role in animal cognition. Philos. Trans. R. Soc. Lond., B, Biol. Sci. 367, 2704–2714.

Shen, K., Hutchison, R.M., Bezgin, G., Everling, S., McIntosh, A.R., 2015. Network structure shapes spontaneous functional connectivity dynamics. J. Neurosci. 35, 5579–5588.

Shine, J.M., Koyejo, O., Bell, P.T., Gorgolewski, K.J., Gilat, M., Poldrack, R.A., 2015. Estimation of dynamic functional connectivity using Multiplication of Temporal Derivatives. NeuroImage 122, 399–407.

Sigman, M., Dehaene, S., 2008. Brain mechanisms of serial and parallel processing during dual-task performance. J. Neurosci. 28, 7585–7598.

Smith, S.M., et al., 2013. Resting-state fMRI in the Human Connectome Project. NeuroImage 80, 144–168.

Sporns, O., Betzel, R.F., 2015. Modular Brain Networks. Annu Rev Psychol 67, annurev–psych–122414–033634.

Steriade, M.M., McCarley, R.W., 2013. Brainstem control of wakefulness and sleep. Springer.

Tognoli, E., Kelso, J., 2014. The metastable brain. Neuron.

Utevsky, A.V., Smith, D.V., Huettel, S.A., 2014. Precuneus is a functional core of the default-mode network. J. Neurosci. 34, 932–940.

van den Heuvel, M.P., Bullmore, E.T., Sporns, O., 2016. Comparative Connectomics. Trends in Cognitive Sciences. doi:10.1016/j.tics.2016.03.001

van den Heuvel, M.P., Sporns, O., 2013. Network hubs in the human brain. Trends in Cognitive Sciences 17, 683–696.

Van der Werf, Y.D., Witter, M.P., Groenewegen, H.J., 2002. The intralaminar and midline nuclei of the thalamus. Anatomical and functional evidence for participation in processes of arousal and awareness. Brain Research Reviews 39, 107–140.

van Ravenzwaaij, D., Oberauer, K., 2009. How to use the diffusion model: Parameter recovery of three methods: EZ, fast-dm, and DMAT. Journal of Mathematical Psychology.

Varela, F., Lachaux, J.P., Rodriguez, E., 2001. The brainweb: phase synchronization and large-scale integration. Nature reviews Neurosci 2, 229–239.

Vatansever, D., Menon, D.K., Manktelow, A.E., Sahakian, B.J., Stamatakis, E.A., 2015. Default Mode Dynamics for Global Functional Integration. J. Neurosci. 35, 15254–15262.

Wagenmakers, E.-J., van der Maas, H.L.J., Grasman, R.P.P.P., 2007. An EZ-diffusion model for response time and accuracy. Psychon Bull Rev 14, 3–22.

Yellin, D., Berkovich-Ohana, A., Malach, R., 2015. Coupling between pupil fluctuations and resting-state fMRI uncovers a slow build-up of antagonistic responses in the human cortex. NeuroImage 106, 414–427.

Zalesky, A., Fornito, A., Cocchi, L., Gollo, L.L., Breakspear, M., 2014. Time-resolved resting-state brain networks. Proceedings of the National Academy of Sciences 111, 10341–10346.